\documentclass[conference]{IEEEtran}
\IEEEoverridecommandlockouts

\usepackage{cite}
\usepackage{amsmath,amssymb,amsfonts}
\usepackage{graphicx}
\usepackage{textcomp}
\usepackage{xcolor}
\usepackage{url}
\usepackage{enumitem}
\usepackage{graphicx}   
\usepackage{subcaption} 
\usepackage{algorithm}
\usepackage{algpseudocode}

\usepackage{multicol}

\usepackage{booktabs}
\usepackage[binary-units=true]{siunitx}
\usepackage{array,multirow}
\usepackage{hyperref}
\usepackage{float}
\usepackage[symbol, hang,flushmargin]{footmisc}
\usepackage{bm}

\newcolumntype{L}[1]{>{\raggedright\let\newline\\\arraybackslash\hspace{0pt}}m{#1}}
\newcolumntype{C}[1]{>{\centering\let\newline\\\arraybackslash\hspace{0pt}}m{#1}}
\newcolumntype{R}[1]{>{\raggedleft\let\newline\\\arraybackslash\hspace{0pt}}m{#1}}

\def\BibTeX{{\rm B\kern-.05em{\sc i\kern-.025em b}\kern-.08em
    T\kern-.1667em\lower.7ex\hbox{E}\kern-.125emX}}
\begin{document}

\newcommand{\secref}[1]{Sec.~\ref{#1}}
\newcommand{\figref}[1]{Fig.~\ref{#1}}
\newcommand{\tabref}[1]{Table~\ref{#1}}

\title{ \LARGE Advancing AI-assisted Hardware Design with  Hierarchical Decentralized Training and Personalized Inference-Time Optimization \\ 
}
\author{
\IEEEauthorblockN{
    Hao (Mark) Chen\IEEEauthorrefmark{2}, 
    Zehuan Zhang\IEEEauthorrefmark{2}, 
    Wanru Zhao\IEEEauthorrefmark{3},
    Nicholas Lane\IEEEauthorrefmark{3}, 
    Hongxiang Fan\IEEEauthorrefmark{2}
}
\IEEEauthorblockA{\IEEEauthorrefmark{2}Imperial College London, London, UK 
\IEEEauthorrefmark{3}University of Cambridge, Cambridge, UK \\
\{hc1620, zehuan.zhang22, w.luk, hongxiang.fan\}@imperial.ac.uk
\{wz341, ndl32\}@cam.ac.uk}
}

\maketitle

\begin{abstract}
Recent years have witnessed a significant increase in the adoption of AI techniques to enhance electronic design automation.
In particular, the emergence of Large Language Models (LLMs) has sparked significant interest in LLM-assisted hardware design generation, spanning applications from classical digital circuits to quantum computing.
Despite substantial progress in this direction,
the quality of LLM-generated hardware design still cannot meet the requirements for practical deployment.
In this work,
we identify three critical challenges hindering the development of LLM-assisted hardware design generation: \textit{1)} limited data availability, \textit{2)} varied data quality, \textit{3)} inadequate inference-time efficiency.  
To address these fundamental challenges,
this paper introduces a two-stage framework for AI-assisted hardware design by exploring decentralized training and personalized inference.
In the first stage, we propose to harness private domain design sources through a hierarchical decentralized training mechanism that addresses data-sharing constraints.
To mitigate the impact of low-quality data, we identify optimization opportunities in hardware generation tasks, using user-defined metrics for model aggregation.
The second stage focuses on client personalization to enhance both speed and quality. We introduce a new metric, Trueput, to analyze    LLM-assisted hardware generation efficiency. To optimize Trueput, we implement personalized inference-time acceleration and customized sampling strategies.
Evaluating both classical and quantum benchmarks,
our experimental results demonstrate that the proposed two-stage framework can significantly improve the model capability for hardware design generation.
As orthogonal enhancements to existing methods, our framework can achieve $33\% \sim 50\%$ semantic accuracy improvement and $2.3$ times speedup, depending on the difficulty of the generation tasks.
Both the code and benchmarks will be released publicly to foster further development in this field.

\end{abstract}


\section{Introduction}\label{sec:intro}
Recent advancements in Large Language Models (LLMs) have demonstrated their great potential in automated software programming~\cite{jiang2024survey} and debugging~\cite{jin2024llms}.
This impressive capability has sparked significant research and industrial interest in leveraging LLMs to automate hardware design for both classical and quantum domains.
In classical hardware design, extensive research has targeted RTL design generation~\cite{thakur2024verigen,chang2023chipgpt,liu2023verilogeval,lu2024rtllm,liu2024rtlcoder}, High-Level Synthesis (HLS) coding~\cite{xu2024optimizing,xiong2024hlspilot,liao2024llms}, and hardware debugging~\cite{fu2023llm4sechw,tsai2023rtlfixer,yan2024assertllm}.
In quantum design generation,
IBM has pioneered the use of LLMs for quantum programming~\cite{dupuis2024qiskit}, which has been integrated into their Qiskit Code Assistant tool\footnote{\url{https://docs.quantum.ibm.com/guides/qiskit-code-assistant}}.

Although significant research efforts have been devoted to exploring LLM-assisted design generation for both classical and quantum hardware, there are still three key challenges hindering their practical use and deployment:
\begin{itemize}[leftmargin=*]
    \item \textit{Challenge-1}: Limited availability of hardware design sources for training. Due to the low-resource nature of hardware description languages, the amount of publicly accessible classical and quantum design sources is much lower than that of software programs. For example, the size of the Qiskit dataset~\cite{dupuis2024qiskit} is more than $1000$ times smaller than that of the Python dataset~\cite{lozhkov2024starcoder}.
    \item \textit{Challenge-2}: Varied quality of training data. High-quality hardware designs are often proprietary and unavailable to the public. The existing training data sourced from public repositories may lack the embedded knowledge necessary for designing high-quality hardware~\cite{liu2024rtlcoder}.
    \item \textit{Challenge-3}: Insufficient generation quality. Existing LLM-assisted methods still exhibit limited accuracy in hardware generation tasks~\cite{lozhkov2024starcoder}. The lack of personalized and customized optimizations during deployment time further limits the potential of LLM for hardware design generation.

\end{itemize}
Therefore, current LLM-assisted hardware generation remains in its early stages, with practical deployment limited.

\begin{figure}
    \centering
    \includegraphics[width=0.99\linewidth]{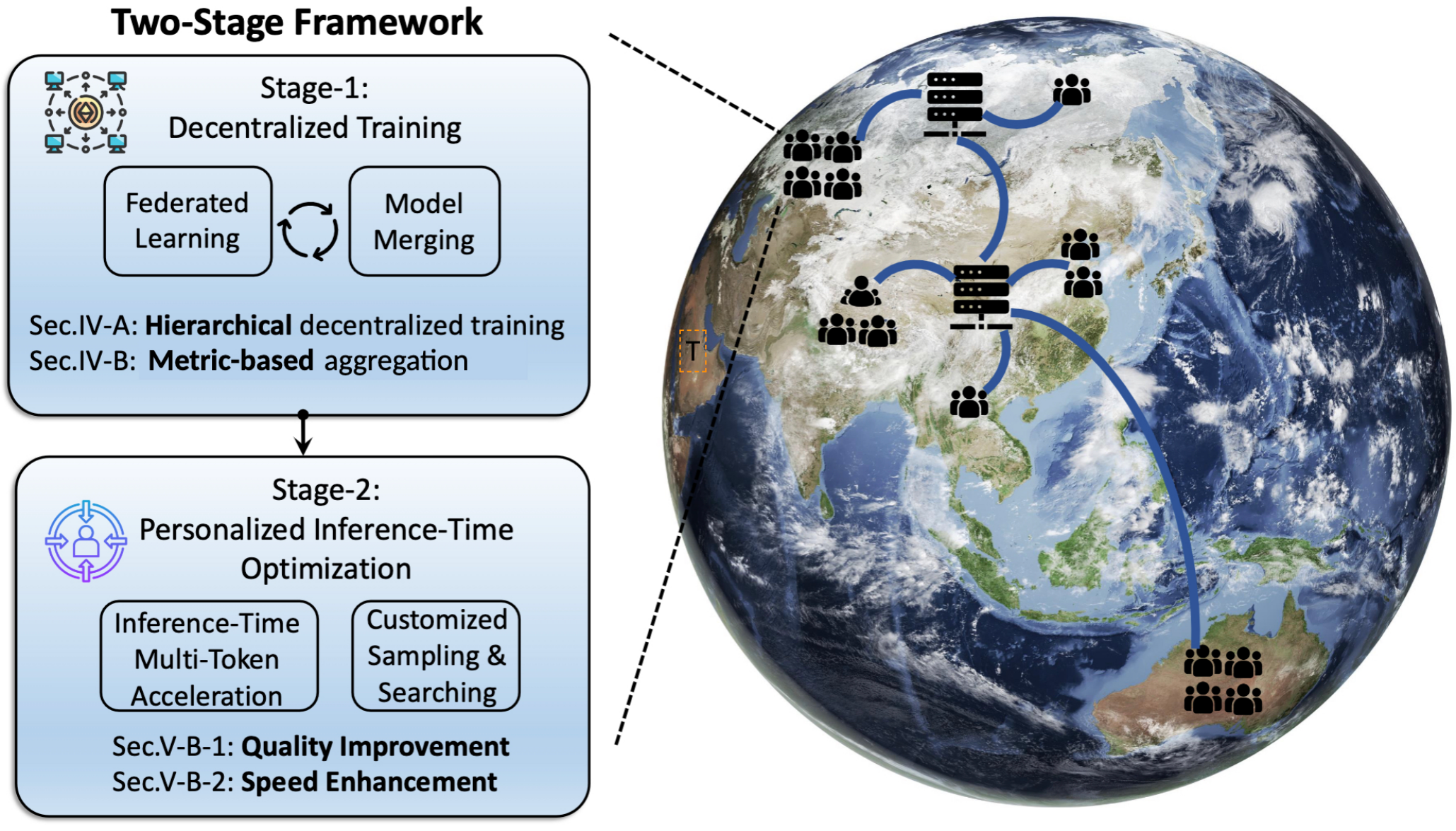}
    \caption{An overview of our proposed two-stage framework for the future of AI-assisted hardware design.
    \vspace{-8mm}
    }
    \label{fig:overview}
\end{figure}

To fully unleash the potential of generative AI for the future of AI-assisted hardware design, this work proposes a two-stage framework by leveraging decentralized and personalized learning.
To address \textit{Challenge-1} of data availability,
we aim to harness private domain design sources through a hierarchical decentralized training mechanism that addresses data-sharing constraints.
This proposed approach includes recent advancements in federated learning~\cite{nguyen2021federated} and model merging~\cite{yang2024model},
introducing a novel hierarchical model update mechanism to facilitate broad adoption among users and organizations by accommodating varied hardware capabilities, communication infrastructures, and individual preferences.
To tackle \textit{Challenge-2} of data quality,
we propose a metric-based model aggregation and merging strategy.
Although data quality control in the decentralized setting is a challenging task~\cite{albalak2024survey, elazar2024whats, kwon2024datainf},
we identify the unique optimization opportunities for hardware design code such as correctness and performance.
To overcome \textit{Challenge-3} of generation quality,
we propose personalized inference-time optimizations to enhance the generation capability.
This includes speed optimization using inference-time multi-token acceleration and quality improvement through customized output token sampling.
As shown in~\figref{fig:overview}, these optimizations follow the decentralized training process, forming a general two-stage framework for the future of AI-assisted hardware design.

Overall, our contributions are summarized as follows:
\begin{itemize}[leftmargin=*]
    \item A hierarchical decentralized training paradigm with metric-based model aggregation, facilitating a broader and more diverse pool of participants for collaborative training in AI-assisted hardware design (\secref{sec:decentralized}).
    \item Personalized inference-time acceleration with customized sampling strategies, improving both efficiency and design quality of LLM for automatic hardware generation (\secref{sec:inf_opt}). 
    \item A comprehensive benchmarking and evaluation of the proposed two-stage framework in both classical and quantum hardware design, highlighting the effectiveness and versatility of our approach (\secref{sec:overview_benchmark} \& \secref{sec:eval}).
\end{itemize}

\section{Background and Related Work}

\subsection{Decentralized Training}

Decentralized training distributes the model training across multiple nodes and devices, with only the communication of weights or gradients for model updates.
The primary benefits of decentralized training are two-fold:
\textit{1)} Proprietary data preservation: By maintaining data locally on the client side for training, decentralized training circumvents data-sharing constraints.
\textit{2)} Compute efficiency: The vast computational resources available on billions of client devices can be utilized for training. 
In this paper, we mainly focus on two mainstreaming decentralized training: federated learning and model merging.
It is worth noting that our proposed framework is general and can be extended to accommodate any decentralized training approach.

\subsubsection{Federated Learning}

As a promising approach to achieve decentralized deep learning~\cite{shokri2015privacy},
federated learning~\cite{mcmahan2017communication} has been extensively studied and optimized over the past decade.
The key concept of federated learning is to move model training from a central server to distributed client devices. 
Depending on the structure of parties (e.g. organizations or individual clients), the scale of participants, and data characteristics, federated learning is typically categorized into cross-device and cross-silo methods~\cite{kairouz2021advances}.
By performing the training locally on client devices, federated learning periodically collects and aggregates model updates from different clients. 
Following the introduction of the classical FedAvg algorithm~\cite{mcmahan2017communication}, recent research in federated learning has primarily focused on addressing challenges related to data and system heterogeneity~\cite{ye2023heterogeneous} to enable practical deployment. 

\subsubsection{Model Merging}
With the recent advancements in language models and the increasing number of open-sourced pre-trained models~\cite{wolf2019huggingface},
significant research efforts have been focused on model merging that integrates the weights of multiple different models to enhance the general capability of the merged model without the need to access the original training data~\cite{ilharco2022editing}.
As the data are not shared during model merging, it provides an efficient and flexible way to learn the different expert knowledge by merging multiple domain-specific models. To preserve the generalizability and capacity of the merged model, various techniques have been introduced such as weighted-based merging~\cite{zhang2024knowledge}, subspace-based methods~\cite{yu2024language}, and routing-based approaches~\cite{muqeeth2023soft}.
More recently, methods designed for efficient and scalable merging of black-box models have also been introduced~\cite{chen2025fw}.

\subsection{Optimization of LLM Inference}
While various techniques have been introduced to improve the inference efficiency of LLM,
this paper mainly focuses on optimization techniques that avoid time-consuming re-training and major modifications to the model architecture.
\subsubsection{Speculative \& Parallel Decoding}
Due to the sequential nature of autoregressive inference, LLMs suffer from data dependency and memory-bound performance. To address this, speculative decoding~\cite{li2024eagle, chengspecinfer} and parallel decoding~\cite{cai2024medusa, chen2024hardware} enable multi-token generation via iterative guess-and-verify strategies.
Speculative decoding uses a separate model to draft multiple tokens, while parallel decoding leverages lightweight prompt tokens and embeddings. In both methods, the original model verifies the generated tokens.
\subsubsection{Inference-Time Scaling Strategy}
As LLM training improvements slow~\cite{snell2024scaling}, recent work has focused on inference-time optimization. One approach is self-refinement, such as recursive introspection~\cite{qu2024recursive}, where the model iteratively improves its answers based on prior outputs. Another is the search-and-verify paradigm~\cite{lightman2023let}, where multiple samples are generated and evaluated by a verifier—either a reward model or end-to-end evaluator—to select the best solution.

\subsection{Related Work}

Hierarchical training has been investigated in previous research on federated learning to address network heterogeneity issues~\cite {abad2020hierarchical, hudson2024flight, gao2024federated}.
\textbf{Distinct from prior approaches},
\textbf{this paper considers diverse clients' conditions in the context of hardware design generation and explores a hierarchical decentralized training with a novel hybrid use of federated learning and model merging}, encouraging a broader and more diverse pool of participants.

Applying client personalization after the training of global models has been investigated in the contexts of federated learning~\cite{kulkarni2020survey}.
Different basic fine-tuning approaches have been employed for model personalization, such as regularised fine-tuning~\cite{t2020personalized} and selective parameter method~\cite{liang2020think}. More advanced techniques, such as meta-learning~\cite {lee2024fedl2p}, have also been explored for client personalization.
\textbf{Unlike previous methods, our framework integrates inference-time acceleration with a novel budget-aware sampling strategy driven by our novel metric Trueput}.

\section{Framework Overview and Benchmarks}\label{sec:overview_benchmark}

\subsection{Framework Overview}

An overview of our proposed framework is illustrated in~\figref{fig:overview}.
Designed to leverage private-domain data for model training with privacy considerations while maximizing deployment efficiency and performance, our framework mainly consists of two stages: decentralized training and personalized inference-time optimization.
These stages can be applied iteratively to collaboratively enhance the model's capabilities for AI-assisted hardware design.

The first stage of decentralized training features a hierarchical mechanism (\secref{subsec:hierarchy}) with hybrid federated learning and model merging, which facilitates a broader and more diverse pool of participants by considering varying hardware capabilities, connection restrictions, and individual preferences.
In the second stage, 
different inference-time optimizations are personalized (\secref{sec:inf_opt}) for each client to enhance the inference speed and generation quality, with different hyperparameters optimized and customized to maximize the deployment performance and efficiency for diverse use cases.

\subsection{Benchmarks}\label{subsec:benchmark}

To demonstrate the effectiveness of our approaches,
we perform evaluation on two different benchmarks: one for classical hardware and the other for quantum hardware. The code files for both scenarios are sourced from publicly available GitHub repositories with compatible licenses, reflecting real-world data heterogeneity.

\textbf{Classical Hardware Benchmark.} 
To validate the applicability of our framework in facilitating classical hardware designs,
we evaluate it on a C-based High-Level Synthesis (HLS) benchmark~\cite{gai2025exploring} comprising 7437 training samples and 1860 test samples. Each sample consists of a high-level design description and a canonical HLS program. The HLS designs include a wide range of domains, such as Matrix and Linear Algebra Operations, Scientific Simulation, etc.


\textbf{Quantum Hardware Benchmark}
To evaluate the effectiveness of our approach in quantum hardware generation, we use a Qiskit benchmark~\cite{campbell2025enhancing} with 10896 training samples and 50 test samples. Each training sample contains a Qiskit program with human-written comments, while each test sample consists of a functionality description and a canonical Qiskit program.


\section{Decentralized Training}\label{sec:decentralized}

\subsection{Hierarchical Decentralized Training}\label{subsec:hierarchy}

Federated learning has demonstrated its potential for decentralized training, with both cross-silo and cross-device settings studied for various user scenarios.
However, its practicality and effectiveness might decrease when deploying it for clients with poor or unreliable communication.
This challenge becomes even more pronounced in extra-large-scale collaborative training settings, where geometric and infrastructural restrictions might affect deployment feasibility.
Additionally, most federated learning approaches assume a shared network architecture for locally trained models, which becomes impractical in the context of LLM due to the high computational and memory requirements for LLM training on client devices.
To promote the broader adoption of decentralized training for foundation models in AI-assisted hardware design, this paper proposes a hierarchical decentralized training scheme.

\begin{figure}
    \centering
    \includegraphics[width=0.99\linewidth]{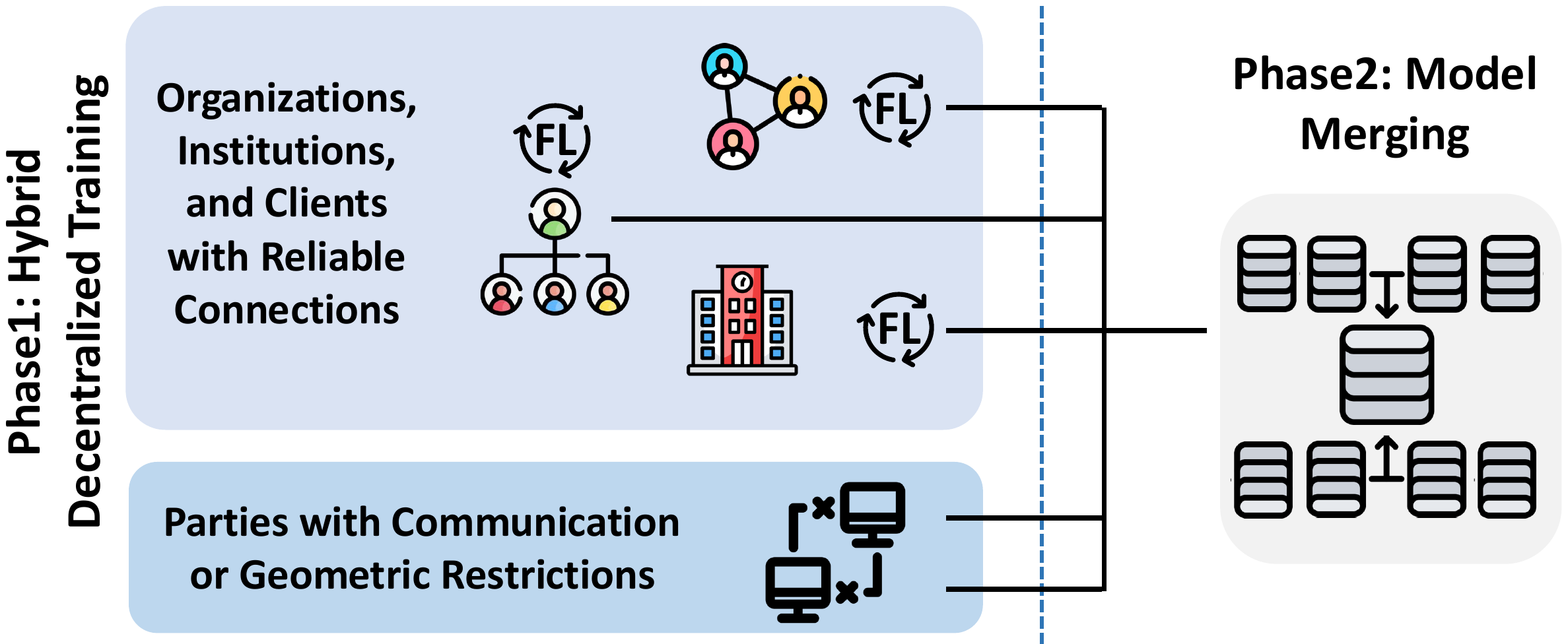}
    \caption{The vision and overview of our proposed framework for the future of AI-assisted hardware design.
    }
    \label{fig:hierachy}
\end{figure}


As illustrated in~\figref{fig:hierachy} and Algorithm~\ref{alg:hierarchical_training}, 
our hierarchical decentralized training consists of two tiers.
The first tier is referred to as hybrid decentralized training. For clients or organizations with reliable communication channels,
federated learning is employed for collaborative training. 
Multiple clients will employ federated learning within each group independently, resulting in several separately trained federated models.
Meanwhile, for parties with isolated environments due to geographical or infrastructural limitations, individual local training is performed.
In the second tier, different models with diverse domain knowledge, learned via either federated or local training,
are combined together using model merging techniques.
This hierarchical, two-tier decentralized training framework enables efficient utilization of private domain data regardless of physical or regulatory restrictions.

\newlength{\textfloatsepsave} 
\setlength{\textfloatsepsave}{\textfloatsep} 
\setlength{\textfloatsep}{0pt} 
\begin{algorithm}
\caption{Hierarchical Decentralized Training}\label{alg:hierarchical_training}
\begin{algorithmic}[1]
\State \textbf{Notation}
\State \( \mathcal{C} \): Set of all clients
\State \( \mathcal{C}_F \): Subset of clients with reliable communication, partitioned into \( G \) groups \( \mathcal{C}_g^F \), where \(g \in \{1, 2, \ldots, G\}\)
\State \( \mathcal{C}_L \): Subset of clients with no reliable communication
\State \( \text{FL}(\cdot) \): Federated learning function 
\State \( \text{LT}(\cdot) \): Local training function
\State \( \text{MM}(\cdot) \): Model merging function
\State \( M_\text{global} \): Final global model after merging
\State \textbf{Tier 1: Hybrid Decentralized Training}
\For{each group \( \mathcal{C}_g^F \) in \( \mathcal{C}_F \)}
    \State Train model \( M_g^F = \text{FL}(\mathcal{C}_g^F) \) \Comment{Federated Learning}
\EndFor
\For{each client \( C_i^L \) in \( \mathcal{C}_L \)}
    \State Train model \( M_i^L = \text{LT}(C_i^L) \) \Comment{Local Training}
\EndFor
\State \textbf{Tier 2: Model Merging}
\State Gather: \( \mathcal{M} = \{M_1^F, \ldots, M_G^F\} \cup \{M_i^L \mid C_i^L \in \mathcal{C}_L\} \)
\State Merge: \( M_\text{global} = \text{MM}(\mathcal{M}) \)
\State \textbf{Output:} Global model \( M_\text{global} \)
\end{algorithmic}
\end{algorithm}

\subsection{Metric-based Aggregation}

Adaptive methods like client selection and quality-aware aggregation~\cite{qi2024model} have been explored in federated learning and model merging, primarily for classification and segmentation tasks. Their effectiveness in generative AI remains underexplored, largely due to challenges in evaluating generated content. Metrics like perplexity depend on reference outputs and often fail to capture functional equivalence—e.g., semantically identical programs with different styles may score differently. Neural metrics such as LLM-as-Judge offer alternatives but often lack explainability and analytical rigor.

To address these challenges, this paper identifies a key optimization opportunity in hardware design generation tasks. Unlike traditional generative tasks, hardware generation inherently provides quantitative evaluation metrics—including design syntax accuracy, hardware functional correctness, and hardware latency—that can serve as robust criteria for model aggregation and merging. Leveraging this observation, we propose a flexible aggregation framework that enables users to define custom metrics for weighting model contributions.
Formally, given the \( i \)-th client model \( M_i \) from a set of \( N \) client models, the final aggregated model \( M_f \) is computed as \(M_f = \sum_{i=1}^{N} g(M_i) \cdot M_i \), where \( g(\cdot) \) is a user-defined metric applied to a client model to determine its contribution.
For instance, \( g(\cdot) \) could be defined using syntax accuracy to filter out model weights from clients trained on syntactically incorrect data, thereby ensuring high-quality training data for the aggregated model.

It is worth noting that our framework is not restricted to using hardware-specific metrics such as syntax accuracy and functional correctness. The framework is designed to accommodate a wide variety of model aggregation strategies, improving the versatility and facilitating broader adoption of our approach. 
For example, if $g(.)$ is parameterized as the ratio of client training samples, the aggregation replicates the FedAvg algorithm. By enabling customization of $g(.)$, our framework caters to diverse requirements across different hardware generation tasks, allowing users to tailor aggregation strategies to their specific needs.

\section{Personalized Inference-Time Optimizations}\label{sec:inf_opt}

\subsection{Trueput: Efficiency Analysis for Design Generation}

To analyze the efficiency of LLM-assisted design generation, we propose a new metric, \textbf{Trueput}, which quantifies the number of functionally correct designs generated per unit of time. It is defined as:
\begin{equation}\label{eq:trueput1}
    \textbf{Trueput} = \frac{\text{Pass@k}}{T_{\text{inf}}}
\end{equation}
where \( \text{Pass@k} \) represents the expected functionality pass rate when \( k \) samples are generated, and \( T_{\text{inf}} \) denotes the expected inference latency per output design.

Next, we analyze Trueput under the constraint of limited computational resources. When batching is employed, the inference latency \( T_{\text{inf}} \) is expressed as \( T_{\text{inf}}(k) \) since the batch size depends on \( k \). According to the Codex~\cite{chen2021evaluating}, an unbiased estimate of \( \text{Pass@k} \) can be written as \(1 - (1 - p)^k \) with functionality pass probability $p$.
Substituting this into the definition of \textbf{Trueput}, we obtain:

\begin{equation}\label{eq:trueput2}
    \textbf{Trueput}_\text{batch} = \frac{1 - (1 - p)^k}{T_{\text{inf}}(k)}
\end{equation}

This formulation enables the analysis of efficiency of the inference framework by accounting for both the accuracy of the generated designs and the latency associated with batching during inference. 
Increasing $\textbf{Trueput}_\text{batch}$ requires decreasing the inference time $T_{inf}(\cdot)$, and improving functionality pass rate $p$.
Given the form in~\eqref{eq:trueput2}, we hypothesize that a global maximum of $\textbf{Trueput}_\text{batch}$ exists at some finite value of $k$, for fixed $p$ and $T_{inf}(\cdot)$. Therefore, the value of $k$ should be optimized for each client to maximize $\textbf{Trueput}_\text{batch}$.
To address these goals, this paper explores personalized test-time optimizations that target both speed enhancement to reduce latency and quality improvement to increase pass rate.

\subsection{Inference-Time Speed and Quality Enhancement}

The scaling law of inference~\footnote{\url{https://openai.com/index/introducing-openai-o1-preview/}} has indicated its potential to improve the performance for most natural language tasks.
In this work, we investigate their effectiveness in hardware design generation and propose customization to further enhance their flexibility and efficiency.

\textbf{Customized Quality Improvement.} Various test-time optimizations~\cite{snell2024scaling} can enhance output generation quality, with popular methods including Best-of-N sampling and beam search. 
Since clients have diverse domains, such as classical or quantum, and their focus on designing different hardware architectures, 
the choice of optimization techniques can vary across different scenarios to maximize the generation quality.
Moreover, these techniques introduce multiple hyperparameters, presenting a design space for optimization.
To leverage this opportunity,  our framework enables clients to customize and optimize their sampling strategy and hyperparameters to meet specific requirements, for example, allowing them to balance hardware design quality and generation latency by adjusting the sampling count. 
\tabref{tab:sampling_strategies} presents the test-time optimization strategies supported in our framework with their associated hyperparameters.
To tailor the optimization strategy for each client, a grid search can be used to tune the optimization configurations at a fixed compute budget.

\begin{table}[ht]
\centering
\caption{Test-Time Optimization Strategies}
\label{tab:sampling_strategies}
\begin{tabular}{C{2. cm}|L{3. cm}|L{3. cm}}
\toprule
\textbf{Sampling}       & \textbf{Description}                   & \textbf{Hyperparameters}                      \\ \hline\hline
Nucleus Sampling        & Selects tokens with cumulative probability $p$ & $p$ (cumulative probability)                 \\ \hline
Temperature Sampling    & Scales token probabilities by temperature     & Temperature, Number of generated candidates  \\ \hline
Top-k Sampling          & Chooses from the top $k$ most probable tokens & $k$ (number of sequences to consider)        \\ \hline
Beam Search             & Expands search using a fixed beam width       & Beam width                                   \\ \bottomrule
\end{tabular}
\end{table}

\textbf{Personalized Inference-Time Acceleration.} Generating an optimized hardware design may require a large number of samples, resulting in high generation latency and energy costs. 
Since the performance of auto-regressive generation in LLM inference is typically memory-bound,
several techniques have been introduced to leverage idle compute resources to accelerate LLM inference.
Among these are speculative decoding~\cite{li2024eagle, chengspecinfer} and parallel decoding~\cite{cai2024medusa, chen2024hardware},
which generate multiple tokens in parallel to improve the processing speed.
However, most existing approaches rely on a separate training process to learn the multi-token generation capability.

In this work, we propose an inference-time learning approach, where each client locally learns acceleration parameters during the model's deployment phase while serving real user requests.
Specifically,
we observe that the learning process of multi-token generation involves tuning the acceleration parameters to approximate the predictive distribution of the original model.
Therefore, rather than depending on a training dataset,
our approach utilizes the generation outputs produced during deployment, while serving user requests, for learning multi-token generation.
This method offers two key benefits.
First, by leveraging user-generated content as labels, the approach can be seen as an unsupervised learning technique, eliminating the need for extra datasets.
Second, the learning process is performed during the model deployment time, avoiding a separate training process to learn multi-token generation.
In this paper, we consider parallel decoding approaches as they are more training-efficient compared to other speculative decoding methods, making it suitable for online learning. The training objective is formulated as follows:

{\scriptsize
\[
\arg\min_\phi \mathbb{E}_{x \sim \mathcal{D}} \left[ 
\mathcal{KL}\big(
P_\text{a}(\mathbf{y}_{t+1:t+k} \mid \mathbf{y}_{1:t}, x; \phi), 
P_\text{o}(\mathbf{y}_{t+1:t+k} \mid \mathbf{y}_{1:t}, x; \theta)
\big) 
\right]
\]
}

where \( \phi \) are the acceleration parameters, and \( \mathcal{D} \) is the deployment data distribution. The \( \mathcal{KL} \)-divergence measures the difference between the \( P_\text{a} \) distribution for acceleration and target \( P_\text{o} \) distributions. \( \mathbf{y}_{t+1:t+k} \) represents the predicted token sequence, and \( \mathbf{y}_{1:t} \) the previously generated tokens by target model with parameter $\theta$.










\section{Experiments}\label{sec:eval}

\begin{figure}
    \centering
    \begin{subfigure}[t]{0.32\linewidth}
        \centering
        \includegraphics[width=\linewidth]{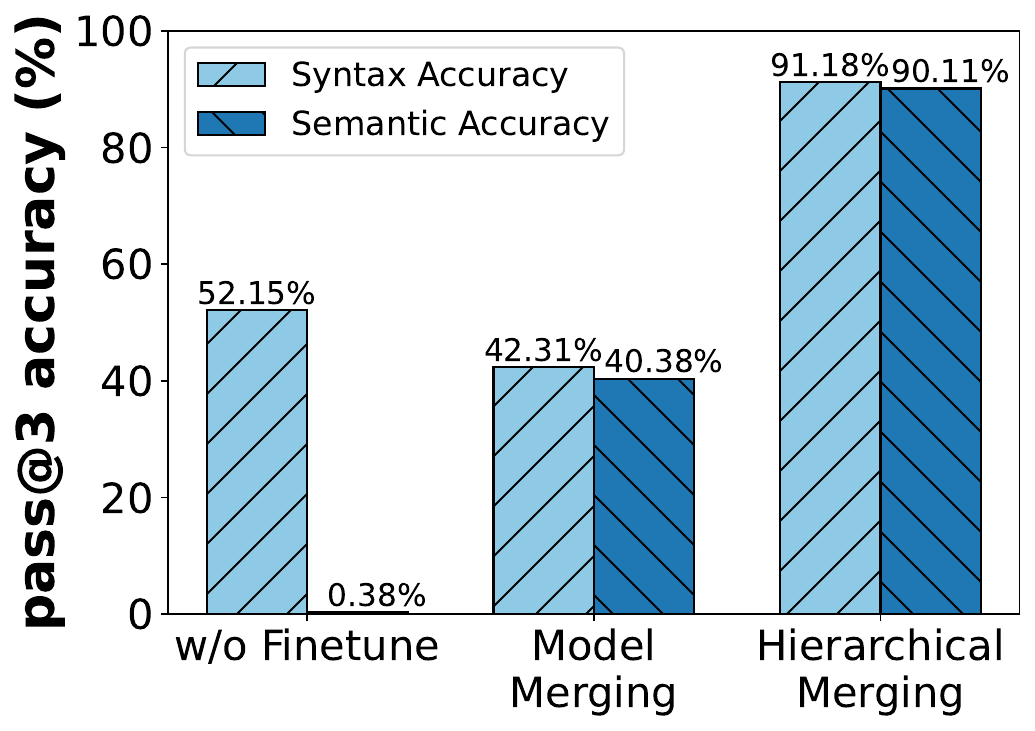}
        \caption{{\footnotesize HLS MachineEval.}}
        \label{fig:hls_machine_hierachy}
    \end{subfigure}
    \hfill
    \begin{subfigure}[t]{0.32\linewidth}
        \centering
        \includegraphics[width=\linewidth]{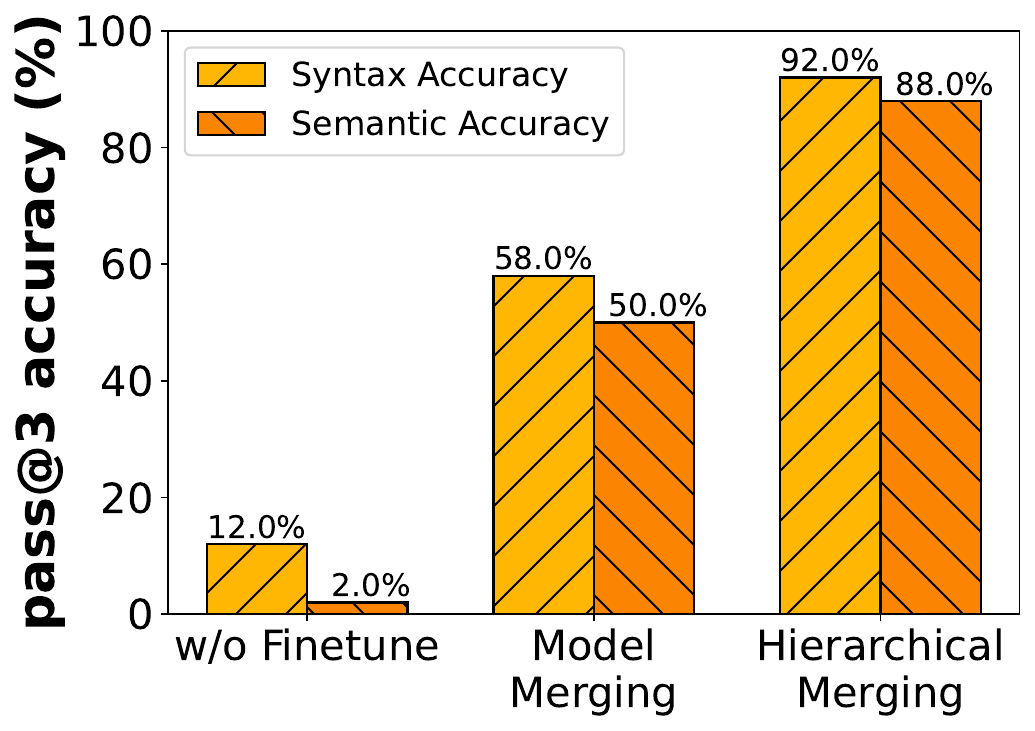}
        \caption{{\footnotesize HLS HumanEval.}}
        \label{fig:hls_human_hierachy}
    \end{subfigure}
    \hfill
    \begin{subfigure}[t]{0.32\linewidth}
        \centering
        \includegraphics[width=\linewidth]{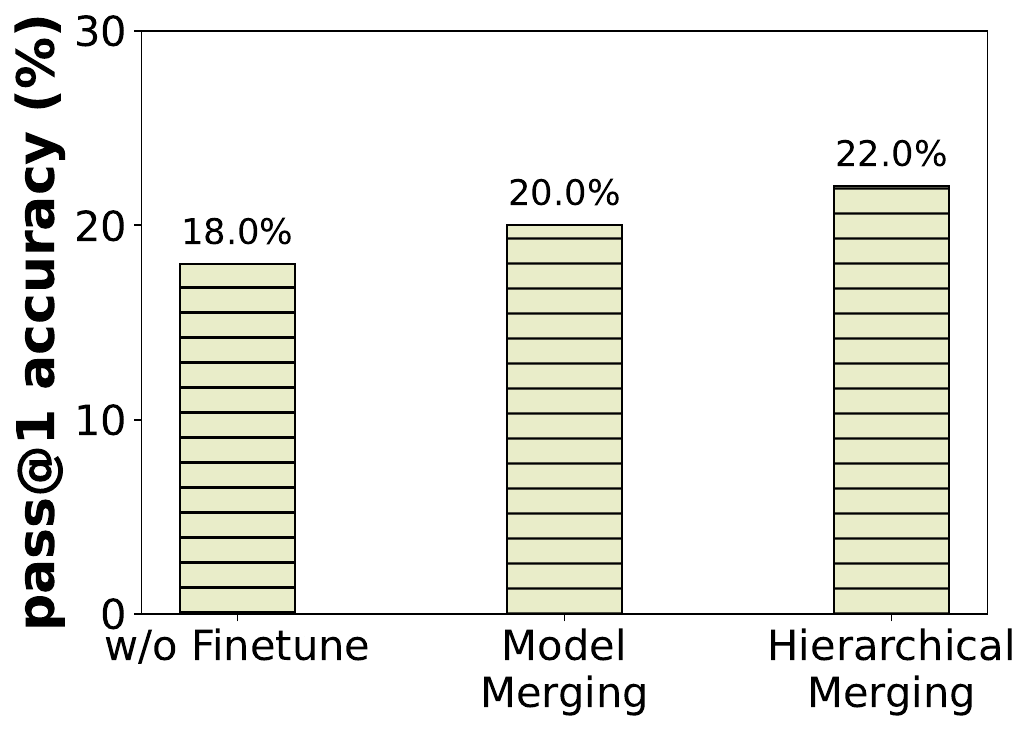}
        \caption{{\footnotesize Qiskit Benchmark.}}
        \label{fig:qiskit_fl_model_merging}
    \end{subfigure}
    \caption{Effect of hierarchical approach on both classical and quantum hardware benchmarks.
    }
    \label{fig:hierachy_combined}
\end{figure}

\begin{figure}
    \centering
    \begin{subfigure}[t]{0.32\linewidth}
        \centering
        \includegraphics[width=\linewidth]{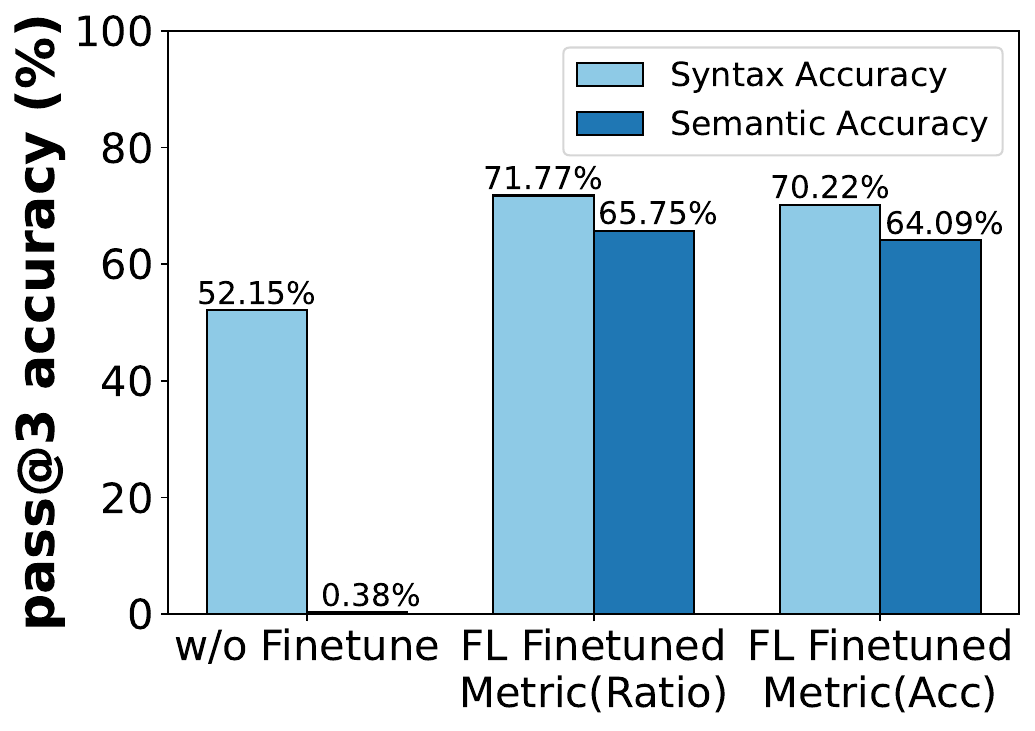}
        \caption{{\footnotesize HLS MachineEval.}}
        \label{fig:hls_machine_fl}
    \end{subfigure}
    \hfill
    \begin{subfigure}[t]{0.32\linewidth}
        \centering
        \includegraphics[width=\linewidth]{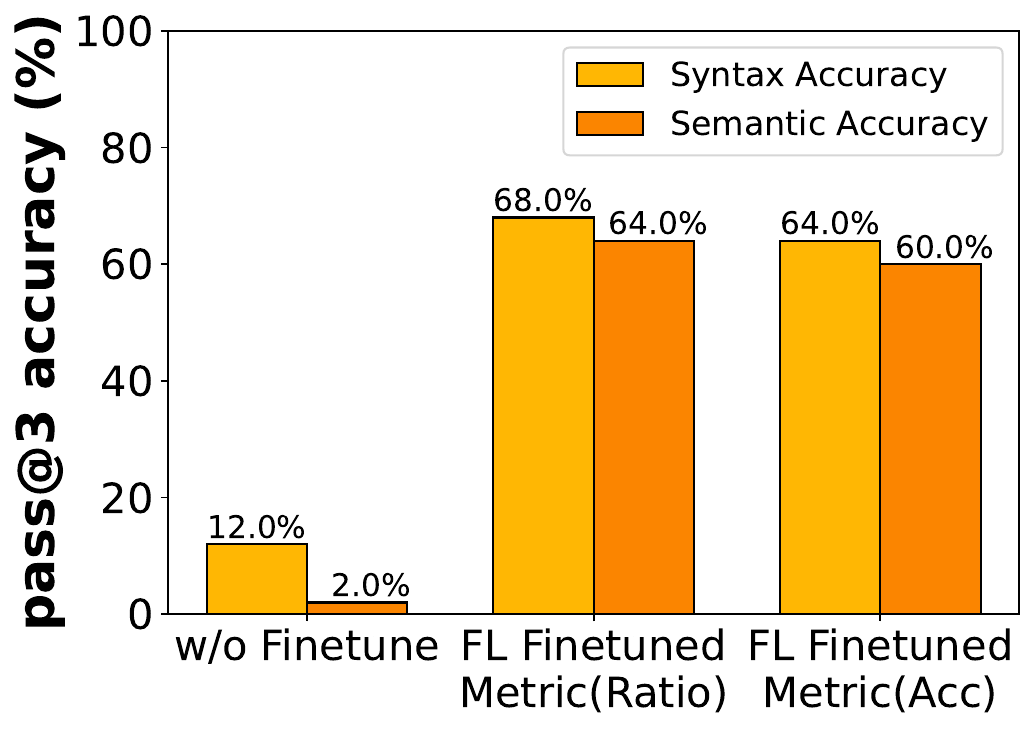}
        \caption{{\footnotesize HLS HumanEval.}}
        \label{fig:hls_human_fl}
    \end{subfigure}
\hfill
    \begin{subfigure}[t]{0.32\linewidth}
        \centering
        \includegraphics[width=\linewidth]{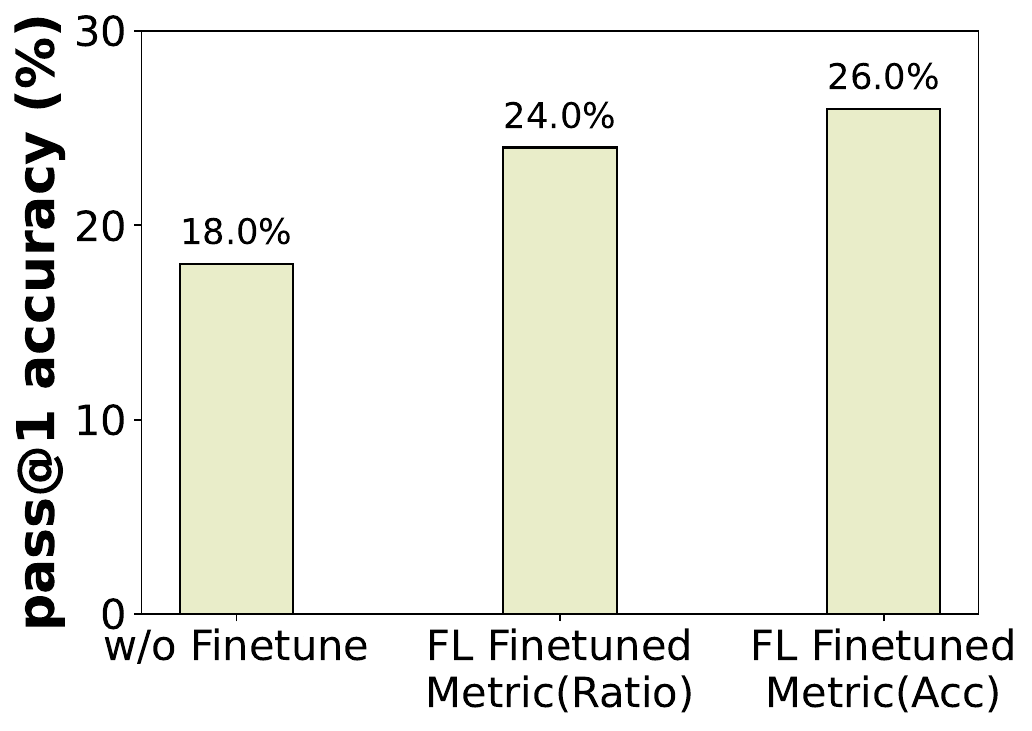}
        \caption{{\footnotesize Qiskit Benchmark.}}
        \label{fig:qiskit_fl}
    \end{subfigure}
    \caption{
    Evaluation of federated learning on both classical and quantum hardware benchmarks.
    }
    \label{fig:fl_combined}
\end{figure}

\subsection{Evaluation Setup}

\textbf{Models and Datasets} Our proposed framework is applicable to a wide range of machine learning methods. However, due to their growing popularity and practical relevance, we focus on LLMs in our experiments. For classical hardware experiments, we use CodeLlama-7B~\cite{roziere2023code} as the base model and the HLS benchmark described in~\secref{subsec:benchmark}.
This benchmark contains machine-generated instructions (MachineEval) produced by GPT for HLS generation.
To evaluate the model's generalizability, we involve human experts to manually refine 50 samples, creating a HumanEval version. 
For Qiskit quantum design generation, we use StarCoder2-3B~\cite{lozhkov2024starcoder} with the dataset introduced in~\secref{subsec:benchmark}.

\textbf{Federated Learning}. For both the classical and quantum benchmarks, we simulate real-world data heterogeneity by training 40 clients on datasets partitioned using a Dirichlet distribution~\cite{li2022federated}, based on the repository IDs of the source code. Each round involves training for one epoch with 10\% of the clients participating. Two aggregation metrics were tested: the number of data samples (Ratio) and hardware syntax accuracy (Acc). A separate validation dataset was used to calculate syntax accuracy.

\textbf{Model Merging}. Hardware syntax accuracy on a validation dataset was used as the weight for model aggregation for both benchmarks. DARE~\cite{yu2024language} was used for hierarchical model aggregation. In this setting, the datasets are partitioned in a manner similar to federated learning, with each client performing local training on its own subset of data.

\textbf{Personalized Test-Time Optimization}. We adopt parallel decoding~\cite{chen2024hardware} for inference-time acceleration. A validation dataset with hardware design instructions is used to simulate user requests. Different sampling strategies and the associated hyperparameters are summarized in~\tabref{tab:sampling_strategies}.

\subsection{Effect of Hierarchical Approach}

To evaluate the effectiveness of our proposed hierarchical approach,
we conduct experiments on both classical and quantum benchmarks, as shown in~\figref{fig:hierachy_combined}. 
We compare our method against two baselines: the base model without fine-tuning and a model merging without hierarchical aggregation.
For classical hardware generated via HLS, we assess both syntax and semantic accuracy with template generation enhancement. For the quantum benchmark, we primarily focus on semantic accuracy evaluation.
In both classical and quantum evaluations,
our approach demonstrated accuracy improvement.
As shown in~\figref{fig:hls_machine_hierachy}\&\ref{fig:hls_human_hierachy},
the hierarchical approach demonstrates significantly greater improvement for classical hardware generation tasks in both MachineEval and HumanEval, achieving nearly an 80\% increase in syntax and semantic accuracy compared to the model without fine-tuning, and approximately 50\% over the model obtained through model merging.
While improvements on the Qiskit benchmark were less pronounced due to the increased complexity of quantum circuit design, our model still performs comparably to the centrally trained baseline~\cite{campbell2025enhancing}. Its performance can be further enhanced by integrating it into the multi-agent framework proposed in~\cite{campbell2025enhancing}, which incorporates Retrieval-Augmented Generation (RAG), Chain-of-Thought (CoT) reasoning, and a semantic analyzer.

\textbf{Training Overhead and Communication Costs}: Our hierarchical approach significantly cuts communication versus standard FL. Standard FL requires $N \times R$ central updates ($N$ clients, $R$ rounds). On the other hand, our hierarchical approach confines the frequent communication within $G$ groups ($N_{FL}$ clients total), with only \textit{one} central merge involving $G$ group models and $N_L$ local models ($N_{FL}+N_L=N$). Thus, central communication drops from $O(N \times R)$ to $O(G+N_L)$ transfers, significantly improving scalability and reducing bandwidth needs, as $G+N_L \ll N \times R$ typically. This structure efficiently leverages clients' idle compute, requiring significantly less compute from central server compared to standard central training.

\subsection{Evaluation of Federated Learning}

\figref{fig:fl_combined} present the evaluation of federated learning on classical and quantum benchmarks using two different aggregation strategies.
As shown in~\figref{fig:qiskit_fl},
leveraging hardware syntax accuracy during model aggregation achieves the best result in the Qiskit Benchmark.
For classical hardware generation as shown in~\figref{fig:hls_machine_fl}\&\ref{fig:hls_human_fl},
Both ratio-based and Acc-based approaches achieve similar results in MachineEval and HumanEval, with up to a 60\% increase in semantic accuracy. 
These findings demonstrate the effectiveness and flexibility of federated learning with metric-based aggregation in training models for both classical and quantum hardware design generation.



\subsection{Personalized Inference-Time Optimizations}

\begin{figure}
    \centering
    \begin{subfigure}[t]{0.48\linewidth}
        \centering
        \includegraphics[width=\linewidth]{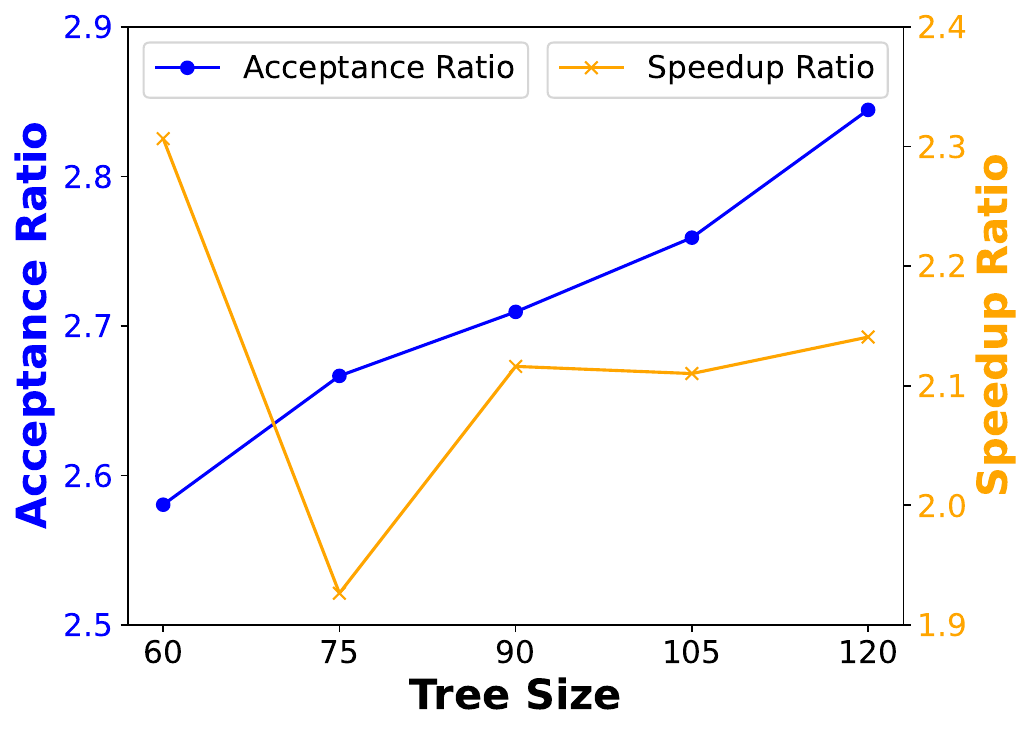}
    \end{subfigure}
    \begin{subfigure}[t]{0.48\linewidth}
        \centering
        \includegraphics[width=\linewidth]{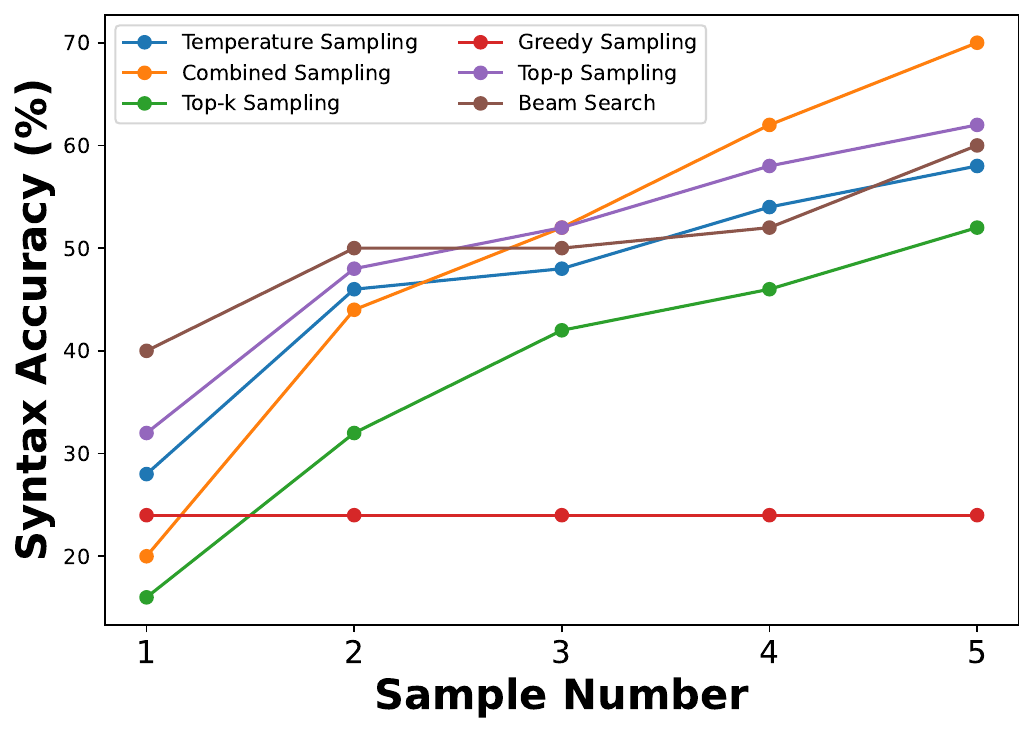}
    \end{subfigure}
    \hfill
    \caption{
    \textbf{Personalized Test-Time Optimization}. Left: multi-token generation. Right: Customized sampling for HLS models. Combined Sampling uses both top-k and top-p filtering.
    \vspace{-3mm}
    }
    \label{fig:acceleration}
\end{figure}

As discussed in~\secref{sec:intro}, the lack of personalized optimizations restricts the potential for maximizing both speed and quality in model inference. 
Thus, we evaluate the impact of two test-time optimizations: multi-token generation for acceleration and customized sampling for quality improvement. 
We show that by tailoring the multi-token generation configuration and sampling strategies to the client’s compute budget, the optimization achieves 2.3 $\times$ speedup ratio and up to 46\% syntax accuracy improvement over default greedy decoding.   
In~\figref{fig:acceleration}, we evaluate the speedup ratio with respect to the tree size, a parameter reflecting the token parallelism. 
While the acceptance ratio increases with larger tree sizes, the speedup ratio peaks at a tree size of 60 due to limited idle compute resources. This aligns with previous research ~\cite{chen2024hardware}, which shows that tree size must be optimized per client to achieve maximum speedup.  
In~\figref{fig:acceleration}, we examine how different sampling strategies affect syntax accuracy. Beam search performs best with sample sizes under 3, while combined sampling outperforms others for larger sample sizes. 
Our findings emphasize the need for personalized inference-time optimizations to advance AI-assisted hardware design. The optimal configuration for each client can be determined offline with a one-time computation, avoiding additional system complexity and compute overhead.

\begin{figure}
    \centering
    \begin{subfigure}[t]{0.48\linewidth}
        \centering
        \includegraphics[width=\linewidth]{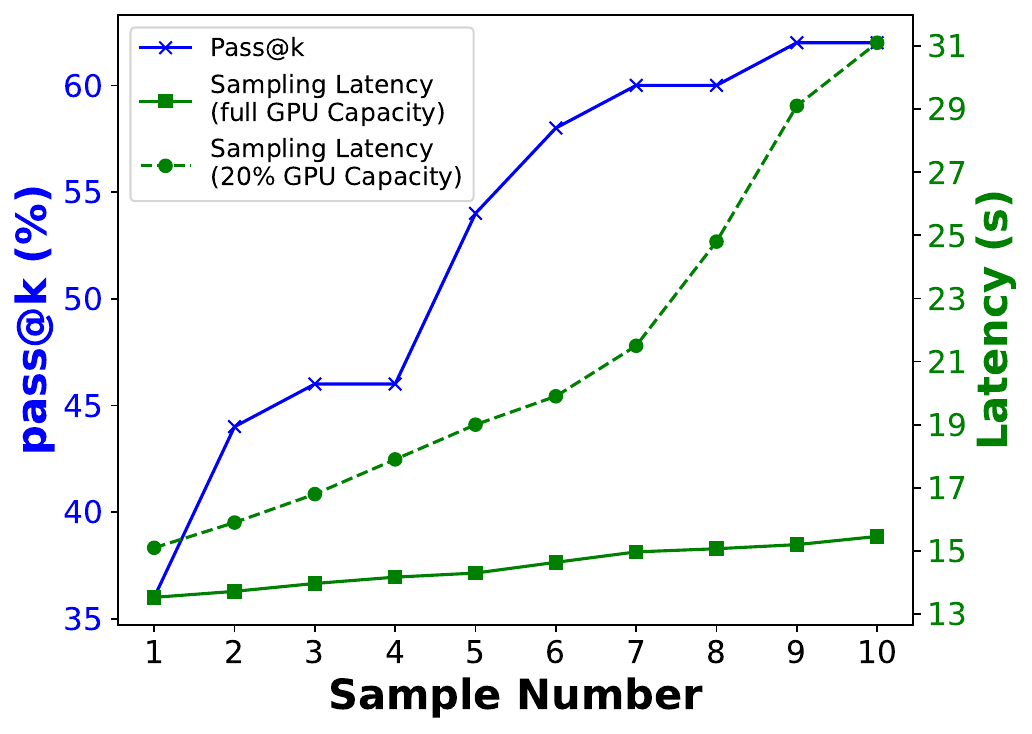}
    \end{subfigure}
    \hfill
    \begin{subfigure}[t]{0.48\linewidth}
        \centering
        \includegraphics[width=\linewidth]{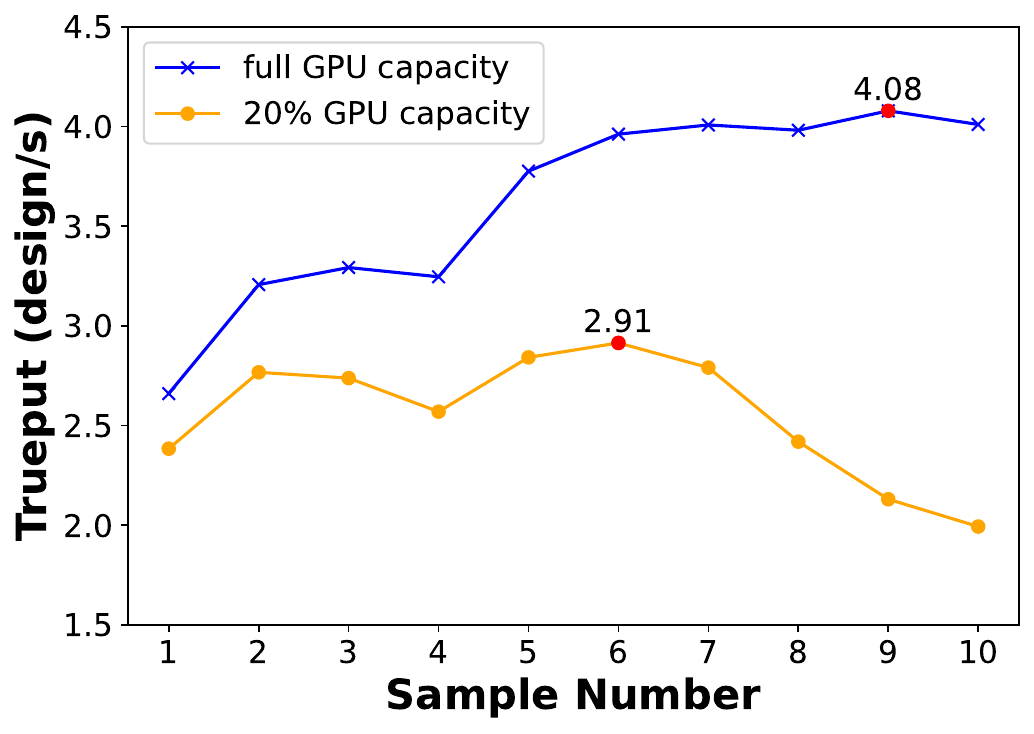}
    \end{subfigure}
    \caption{
        \textbf{Trueput Evaluation}.  
        Left: Pass@k and latency for different sample sizes.  
        Right: \textbf{Trueput} across GPU capacities.
    }
    \label{fig:sampling_latency}
\end{figure}

\subsection{Effect of Sample Number on Trueput Optimization}

In \secref{sec:inf_opt}, we introduce \textbf{Trueput} to analyze the efficiency of design generation. To maximize $\textbf{Trueput}_\text{batch}$, in addition to the previously mentioned personalized test-time optimization, it is important to search for the optimal sample number $k$, as shown in \eqref{eq:trueput2}. As depicted in \figref{fig:sampling_latency}, both Pass@k and sampling latency increase with the sample number $k$, but with different rates. \figref{fig:sampling_latency} shows that a local maximum of $\textbf{Trueput}_\text{batch}$ exists, and the optimal value of $k$ varies depending on the GPU capacity. This observation highlights that the sample number should be optimized per client to achieve the maximum \textbf{Trueput}.

\section{Conclusion}
Recent advancements in AI have shown great potential in revolutionizing the traditional hardware design process.
However,
the limited quality and quantity of available data remain critical barriers to the development of AI-assisted hardware design.
In this paper, the authors argue that addressing this fundamental challenge requires decentralized and personalized learning approaches. To this end, we present a two-stage framework featuring a novel hierarchical decentralized training paradigm with metric-based model aggregation for model training, along with personalized inference-time optimizations to enhance deployment efficiency.
Comprehensive evaluations on both classical and quantum hardware design tasks demonstrate the effectiveness of our approach. We hope that the benchmarking results of this work will encourage broader engagement from both industrial and individual parties in jointly advancing AI-assisted hardware design.
\newpage

\bibliographystyle{plain}
\bibliography{reference}

\end{document}